\documentclass[aps,prb, groupedaddress,twocolumn,longbibliography]{revtex4-1}

\usepackage{graphicx, subfigure}
\usepackage{amsmath}
\usepackage{amsfonts}
\usepackage{amssymb}
\usepackage{bbm}

\usepackage{subfigure}

\usepackage{color}

\begin{document}


\title{Fast and selective phonon-assisted state preparation of a quantum dot \\ by adiabatic undressing}

\author{A.M.~Barth$^{1}$, S.~L\"uker$^{2}$, A.~Vagov$^{1}$, D.E.~Reiter$^{2,3}$, T.~Kuhn$^{2}$ and V.M.~Axt$^{1}$}
\affiliation{$^{1}$Institut f\"{u}r Theoretische Physik III, Universit\"{a}t Bayreuth, 95440 Bayreuth, Germany,}
\email[]{andreas.barth@uni-bayreuth.de}

\affiliation{$^{2}$Institut f\"ur Festk\"orpertheorie, Universit\"at M\"unster, 48149 M\"{u}nster, Germany}

\affiliation{$^{3}$Imperial College London, South Kensington Campus, London SW7 2AZ, UK.}

\date{\today}

\begin{abstract}

We investigate theoretically the temporal behavior of a quantum dot under off-resonant
optical excitation targeted at fast acoustic phonon-assisted state preparation. 
We demonstrate that in a preparation process driven by short laser pulses
three processes can be identified: a dressing of the states during the switch on of the laser pulse,
a subsequent phonon-induced relaxation and an undressing at the end of the pulse.
By analyzing excitation scenarios with different pulse shapes we highlight the decisive
impact of an adiabatic undressing on the final state in short pulse protocols. 
Furthermore, we show that in exciton-biexciton systems the laser characteristics such as
the pulse detuning and the pulse length as well as the biexciton binding energy can be used
to select the targeted quantum dot state.
\end{abstract}

\maketitle


\section{Introduction}

Many of today's proposals for quantum information applications\cite{bouwmeester:00} rely on the controlled 
and fast manipulation of the discrete states of the corresponding devices' underlying
structures. Semiconductor quantum dots (QDs) are frequently discussed as building blocks
for such materials because they hold out the prospect of tailor-made energy spectra
and a high integrability in a solid-state environment. The excitonic QD states
are promising candidates to be used as qubits for quantum computing\cite{biolatti:00, chen:01, li:03, piermarocchi:02, boyle:08, nielsen:00},
while the radiative decay from the biexciton cascade offers the possibility of an on-demand creation of indistinguishable entangled photon 
pairs\cite{moreau:01, stevenson:06, mueller:14}. 

It has been shown that both exciton and biexciton states of a QD can be prepared by
using ultra-fast laser pulses under a variety of excitation conditions\cite{reiter:14,ramsay:10c}.
The most commonly known schemes for this purpose are resonant excitation leading to Rabi rotations\cite{zrenner:02,machnikowski:04, vagov:07, ramsay:10},
different protocols using chirped laser pulses exploiting the adiabatic rapid passage effect \cite{schmidgall:10,simon:11,wei:14,lueker:12, glaessl:13, gawarecki:12, debnath:12, debnath:13, mathew:14},
and phonon-assisted off-resonant driving \cite{glaessl:11b, reiter:12, hughes:13, glaessl:13b}.
Recently, the latter method has also been experimentally demonstrated\cite{ardelt:14, quilter:15, bounouar:15}. 
Indeed, there is an increased interest in this approach,
because it is not only stable against
fluctuations of the applied field intensity, but also leaves the quantum dot transition laser-free, which can be
important when the emitted photons need to be spectrally separated from the laser pulse.
Furthermore, in contrast to the other two protocols the phonon-assisted scheme makes active use of the phonon coupling
and even works the better the stronger this coupling is. 

In this paper, we examine the influence of the pulse shape on the phonon-assisted state preparation.
We identify three distinct processes that take place during the laser-driven evolution of the QD states.
When the pulse starts the coupling to the laser-field leads to a \emph{dressing} of the bare QD states.
This enables a \emph{phonon-induced relaxation} between the dressed states\cite{leggett:87, dekker:87, glaessl:11b}
and finally when the pulse is switched off an \emph{undressing} takes place.
While previous studies mostly concentrated on the phonon induced relaxation and the resulting exciton and biexciton
occupation obtained when applying an off-resonant driving pulse\cite{glaessl:13b}, here we will explain the impact of all three
different processes and examine in detail the role of the switch-on and switch-off phase of the excitation.
We will show that within the relaxation process there is a trade-off situation between a sufficiently fast 
preparation and an optimal preparation fidelity and that for a high fidelity 
preparation of the bare QD states an adiabatic undressing, that can be realized by a long enough switch-off
time, is indispensable in short pulse protocols.
Our analysis also shows that even though Gaussian pulses, as used in very recent experiments\cite{ardelt:14, quilter:15, bounouar:15},
are not the ideal choice for the phonon-assisted protocol they fulfill the requirements for fast state preparation
surprisingly well for a wide intensity range provided that 
the regime of non-adiabatic dynamics is not yet reached. The latter condition sets a lower bound to the pulse
duration.

The pulse characteristics not only allow to control the fidelity of the achieved inversion,
but in addition can be used to select the QD state that is targeted by the phonon-assisted process.
For the two-level case the sign of the pulse detuning determines whether
the QD is driven towards the ground-state or the exciton state. For the exciton-biexciton system
also the biexciton binding energy and the pulse length play a critical role in determining the targeted QD state.
Understanding that the preparation is a three-step process gives us an intuitive answer to the important question which state is selected
by the phonon-assisted preparation scheme.


\section{Model}\label{model}  

\begin{figure}[t]
 \includegraphics[width=\columnwidth]{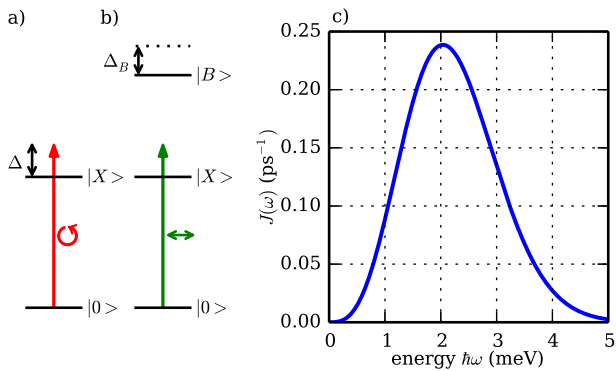}
    \caption{(Color online) Energetic level diagram of (a) the exciton system with circular polarized excitation and (b) the exciton-biexciton system with linear polarized excitation.
    (c) Spectral density of the phonon coupling as a function of energy for the parameters used in our simulations (see text).}
 \label{fig1}
\end{figure}

We consider a strongly confined GaAs QD driven by an external laser field and 
coupled to a continuum of acoustic bulk phonons. Our model is defined by the Hamiltonian
$H = H_{\rm{dot,\,light}} + H_{\rm{dot-ph}}$, i.e., the electronic system coupled
to the laser field and an additional phonon part.
Let us first concentrate on 
\begin{align}
&H_{\rm{dot,\,light}} \!= \sum_\nu \hbar\omega_\nu |\nu\rangle\langle\nu|
                      \!+ \sum_{\nu\nu'} \hbar M_{\nu\nu'}|\nu\rangle\langle\nu'|,
\label{eq_H_dotlight}
\end{align}
where $|\nu\rangle$ are the electronic basis states with corresponding energies $\hbar \omega_{\nu}$. 
The matrix element $M_{\nu\nu'}$ describes the coupling between the QD and the classical laser field using
the common dipole and rotating wave approximations.
In the first part of the paper we will restrict ourselves to a two-level system consisting of the ground-state $|0\rangle$,
for which we set $\hbar\omega_0=0$, and a single exciton state $|X\rangle$ with energy $\hbar\omega_X$ as illustrated in Fig.~\ref{fig1}(a).
The two-level approximation is valid when considering circularly polarized light with a single polarization orientation
and the exchange interaction is negligibly small. The exchange interaction strongly depends on the QD geometry\cite{langbein:04} and 
can be close to zero, as it is favorable for, e.g., entangled photon creation\cite{stevenson:06}.
In this case the non-zero matrix elements of the light-matter coupling are given by
\begin{align}
&M_{0X} = \frac{1}{2}f(t) e^{i\omega_L t}\,, \qquad M_{X0}=M_{0X}^*\, ,
\label{eq_M2LS}
\end{align}
where $f(t)$ is a real pulse envelope function, which in the following is referred to as field strength.
The field strength $f(t)$ is related to the electric field ${\bf E}(t)$ with frequency $\omega_L$ and the QD dipole matrix element
${\bf d}$ by $-{\bf d}\cdot{\bf E}(t)=\frac{\hbar}{2}f(t)e^{-i\omega_L t}$. 

A very useful picture for strongly driven few-level systems, which we will employ to analyze our results, is the dressed state picture. 
The dressed states are the eigenstates of the coupled light-matter Hamiltonian
in a frame rotating with the laser frequency \cite{tannor:07}.
For the two-level system driven by a laser-field with a fixed field strength $f$ and a detuning $\Delta$ they are given by the expressions
\begin{subequations} \label{dressed_states}
\begin{eqnarray}
 |\psi_{\rm{up}}\rangle &=& +\cos(\Theta)|0\rangle + \sin(\Theta)|X\rangle\\
 |\psi_{\rm{low}}\rangle &=& -\sin(\Theta)|0\rangle + \cos(\Theta)|X\rangle
\end{eqnarray}
\end{subequations}
where $\Theta$ is the mixing angle defined by $\tan(\Theta)=\frac{\hbar f}{\Delta+ \hbar\Omega}$ and $\Omega$ is the Rabi frequency given by 
\begin{align}
\hbar \Omega = \sqrt{(\hbar f)^2+\Delta^2}.
 \label{Omega}
\end{align}
The corresponding dressed state energies read
\begin{align}
 E_{\rm{up/low}}=\frac{1}{2}(-\Delta\pm\hbar\Omega).
 \label{dressed_state_energies}
\end{align}
It is worth noting that the contributions of the ground and exciton states to the dressed states vary depending on the
detuning and the field strength.

The final part of this paper will be devoted to excitations with linearly polarized light. In this case one also has to take into account
the biexciton state and the system described so far needs to be extended to a three-level model consisting
of the ground state $|0\rangle$, the single exciton state $|X\rangle$ and the biexciton state $|B\rangle$. 
Note that the single exciton state $|X\rangle$ in the three-level system has different polarization than in the two-level system described further above.
As is illustrated in Fig.~\ref{fig1}(b), the biexciton state has the energy $\hbar\omega_B=2 \hbar\omega_X- \Delta_B$,
where $\Delta_B$ is the biexciton binding energy. 
In the exciton-biexciton system, the non-vanishing dipole matrix elements are given by
\begin{align}
&M_{0X} = M_{XB}= \frac{1}{2}f(t)e^{i\omega_L t}, M_{X0}=M_{BX}=M_{0X}^*.
\label{eq_M3LS}
\end{align}

Let us now focus on the coupling of the QD to the phonon environment. We model the electron-phonon interaction 
by a pure-dephasing Hamiltonian which, together with the free phonon Hamiltonian, has the form
\begin{align}
&H_{\rm{dot-ph}} \!= \! \sum_{\bf q} \hbar\omega_{\bf q}\,b^\dag_{\bf q} b_{\bf q} 
                       + \sum_{{\bf q} \nu} \hbar n_\nu \big(
                      \gamma_{\bf q} b_{\bf q} + \gamma^{*}_{\bf q} b^\dag_{\bf q}
\big)
|\nu\rangle\langle\nu|.
\label{eq_H_phonon}
\end{align}
The operators $b^{\dagger}_{\bf q}$ ($b_{\bf q}$) create (annihilate) a phonon with wave vector ${\bf q}$ 
and linear dispersion $\hbar \omega_{{\bf q}}=\hbar c_s |{\bf q}|$, where $c_s$ denotes the longitudinal sound velocity.
$n_\nu$ counts the number of excitons present in the state $|\nu \rangle$. The coupling constants $\gamma_{\bf q}$ are 
chosen specific for the deformation potential coupling to longitudinal acoustic phonons, which has been shown to be dominant 
for typical self-assembled GaAs QDs\cite{ramsay:10, krummheuer:02}. As described in more detail in Ref.~\onlinecite{krummheuer:02},
the coupling constants $\gamma_{\bf q}$ depend on the electronic wave functions $\Psi_{e(h)}$ as well as on the deformation potential
constants $D_{e(h)}$ for electrons (e) and holes (h), respectively. For simplicity, we assume the wave functions to be the ground-state
solutions of a spherically symmetric harmonic potential, i.e., $\Psi_{e(h)}({\bf r}) \propto \exp\big(-r^2/(2a_{e(h)}^2)\big)$,
and refer to  $a_e$ as the QD radius. The characteristics of the exciton-phonon coupling can be expressed by the spectral density
$J(\omega)=\sum_{\bf q}|\gamma_{\bf q}|^2\delta(\omega-\omega_{\bf q})$, which under the above assumptions reads\cite{krummheuer:02}
\begin{align}
&J(\omega ) =\frac{\omega^{3}}{4\pi^{2}\rho \hbar c_{s}^{5}} 
  \big[D_{e}e^{-\omega^{2}a_{e}^{2}/(4c_{s}^{2})}   
   -   D_{h}e^{-\omega^{2}a_{h}^{2}/(4c_{s}^{2})} \big]^{2}
\label{eq_J}
\end{align}
with $\rho$ being the mass density. For our present calculations we use material parameters specific for GaAs taken from the literature\cite{krummheuer:05}
$\rho\!=\!5370$~kg/m$^3$, $c_s\!=\!5110$~m/s, $D_e=7.0$~eV, and $D_h=-3.5$~eV. For the QD size we choose $a_e=3$~nm and set $a_e/a_h=1.15$ assuming equal potentials 
but taking into account the different effective masses of electrons and holes for GaAs. 
The spectral density for these parameters is shown in Fig.~\ref{fig1}(c) where 
it can be seen that $J(\omega)$ exposes a clear maximum at $\hbar\omega = E_J^{\rm max} \approx 2$~meV. 
The electron-phonon interaction leads to a polaron-shift $\delta_{ph}$ of the exciton and biexciton energy, such that when driving the exciton-to-ground state transition
the resonant excitation is on the shifted exciton energy $\hbar \tilde{\omega}_X = \hbar \omega_X - \delta_{ph}$. In this paper we are mostly interested
in detuned excitations and define the detuning as the difference between the laser energy and the polaron-shifted exciton energy, i.e., 
\[ \Delta = \hbar \omega_L - \hbar \widetilde{\omega}_X. \]

To study the time evolution of the electronic QD occupations under excitation with the laser field we employ an implementation
of a real-time path-integral approach\cite{vagov:11}. This method allows a numerically exact treatment of the above model
despite the infinite number of LA phonon modes and yields the dynamics of the reduced electronic density matrix of the 
QD including arbitrary multi-phonon processes and taking into account all non-Markovian effects. 
We assume the QD to be initially in a product state of the electronic ground-state and a thermal equilibrium of
the phonon modes at temperature $T=1$~K.


\section{Results}\label{results}  

We start by analyzing the phonon-induced relaxation using continuous excitation that is switched on 
instantaneously. Then we will apply short pulses with different pulse shapes to
analyze the influence of the adiabaticity of the dressing and undressing process 
in view of high-fidelity state preparation.
For this analysis we will also visualize the system trajectory on the Bloch sphere.
Finally, we will demonstrate the selective addressing of all three states in the exciton-biexciton
system using the phonon-assisted state preparation protocol.

\subsection{Phonon-induced relaxation}\label{phonon_induced_relaxation}

\begin{figure}[ttt]
 \includegraphics[width=8.7cm]{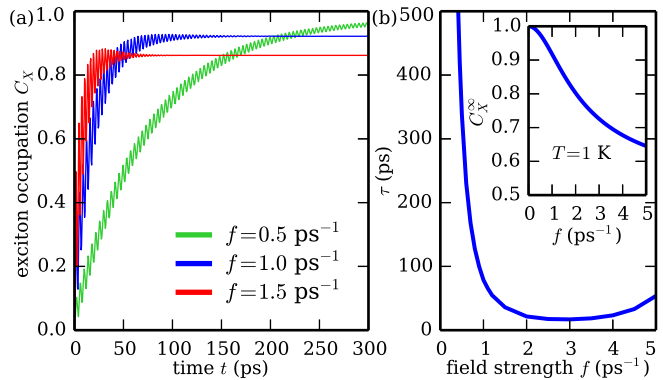}
    \caption{(Color online)
            (a) Exciton occupation $C_X$ as a function of time for different
	    field strengths $f=0.5$~ps$^{-1}$, $1.0$~ps$^{-1}$ and $1.5$~ps$^{-1}$. (b) Time $\tau$ after which
	    the time average of the oscillations of $C_X$ has
	    reached 99\% of $C_{X}^{\infty}$ (see text) as a function of the field strength $f$.
	    Inset of (b): $C_{X}^{\infty}$ as a function of the field strength $f$.
	    The detuning is $\Delta=1.0$~meV.
            }
 \label{fig2}
\end{figure}

The phonon-induced relaxation can be best analyzed by considering a continuous excitation of the QD with a 
constant field strength. Further, the laser field shall be circularly polarized such that our model
can be restricted to two electronic levels, as we explained in the previous section.
Generally, for a weak coupling between the electronic states and the phonon environment the standard expectation
for the driven QD dynamics is that the Markovian approximation is 
well justified and that in the long-time limit the relaxation leads to a thermal occupation
of the dot-photon dressed states\cite{leggett:87, weiss:99} [c.f. Eq.~(\ref{dressed_states})].
In a very good approximation, this has been shown to hold also true for the two-level model of the QD with standard GaAs-type parameters
considered here\cite{glaessl:11b}.
For very low temperatures, the system ends up mainly in the lower dressed state $|\psi_{\rm{low}}\rangle$,
which corresponds to an exciton occupation of
\begin{align}
 C_{X}^{\infty} \approx \cos^2(\Theta)= \frac{1}{2} \left(1 + \frac{\Delta}{\sqrt{\hbar^2 f^2 + \Delta^2}}  \right).
 \label{CX_infty}
\end{align}
Figure~\ref{fig2}(a) shows the simulated temporal evolution of the exciton state under constant excitation for a detuning of $\Delta = 1.0$~meV
and three different field strengths $f$. Here, the laser field is switched on instantaneously at $t = 0$~ps.
The occupations perform damped Rabi oscillations around a mean value that approaches a constant value.
For a decreasing field strength the stationary exciton occupation rises as can also be seen in the inset of Fig.~\ref{fig2}(b)
where $C_{X}^{\infty}$ given by Eq.~(\ref{CX_infty}) is shown as a function of $f$ for $\Delta = 1.0$~meV.
Most importantly for phonon-assisted exciton preparation, the stationary exciton occupation is very close to one and only limited
by the finite temperature for sufficiently small field strengths. Larger values of $f$, however, strongly reduce $C_{X}^{\infty}$.

On the other hand, when we look at the time required to reach the final state, we find that for small field strengths $f$, 
the time to reach the final state becomes longer. This is quantified in Fig.~\ref{fig2}(b), where we have plotted the time $\tau$ it
takes for the mean value of the oscillations to reach 99\% of the exciton occupation expected for a thermal distribution of the
dot-photon dressed states $C_{X}^{\infty}$ as a function of the field strength.
For example at $f = 0.5$~ps$^{-1}$ a relaxation time of several hundred ps is required, which might exceed the time until other
relaxation processes occurring on longer time scales than the phonon-induced relaxation such as the radiative decay
which is not considered here, come into play. Therefore such a low driving strength does not yield a sufficiently fast relaxation
for phonon-assisted state preparation. Indeed, we find that for $f\rightarrow 0 $ the time needed for the state preparation diverges.

To understand the varying relaxation times for different driving strengths one has to keep in mind that the pure dephasing type
phonon coupling does not induce direct transitions between the electronic states, but the phonon-induced relaxation is only
enabled by the laser field. Transitions between the QD states only take place due to a non-vanishing overlap of both of the photon
dressed states with the exciton state, which in turn is coupled to the phonon-environment. 
For an efficient relaxation also the Rabi splitting $\Omega$, i.e. the difference between the two dressed state energies, needs to be 
close to typical phonon energies and ideally matches the maximum of the spectral density of the phonon coupling $J(\omega)$ 
[c.f. Eq.~(\ref{eq_J})]. Both of these properties of the relaxation rate are also captured by a simple application of Fermi's Golden Rule,
which yields a relaxation time between the upper dressed state without phonons and the lower dressed state with one phonon of
\begin{align}
\tau_{\rm relax} = \left(\frac{\pi}{2}\sin^2(2\Theta)\,J(\Omega)\right)^{-1}.
\label{golden_rule}
\end{align}
Here it can be seen that the mixing angle $\Theta$ as well as the spectral density evaluated at the Rabi splitting $J(\Omega)$ enter
the relaxation rate. According to the simulations the maximal relaxation for the detuning $\Delta = 1.0$~meV used here occurs for a field strength around $2.7$~ps$^{-1}$,
which is reflected by the position of the minimum of the time needed for an almost complete relaxation plotted in Fig.~\ref{fig2}(b).
In a good approximation this is expected from the rough estimation in Eq.~(\ref{golden_rule}).
Despite the short relaxation time at the optimal field strength, which is below $20$~ps, such a strong driving field is not applicable for high-fidelity
state preparation either, because the achieved final occupation of below $0.8$ is too far from the desired fidelity of one.

Looking at our results, it becomes clear that the maximum exciton occupation for a given detuning can only be realized for almost
vanishing field strengths where on the other hand it takes arbitrarily long times to complete the relaxation process.
Therefore, we conclude, that when only the phonon relaxation process of the preparation takes place, 
there is a trade-off between reaching the targeted state with high fidelity and realizing fast preparation times.

\subsection{Dressing and Undressing}\label{dressing_undressing}

\begin{figure}[t]
 \includegraphics[width=8.7cm]{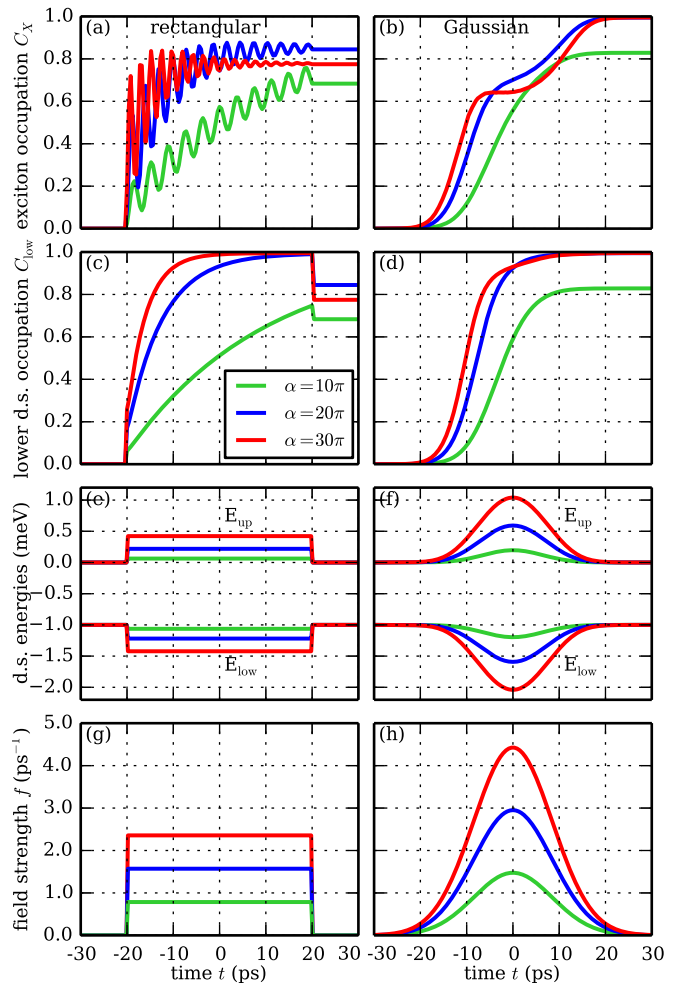}
    \caption{(Color online)
         Dynamics for an off-resonant excitation of the QD with rectangular pulses (left) having a length of $40$~ps and Gaussian pulses (right) 
         with a FWHM of $20$~ps. The pulse areas are $\alpha=10\pi$ (green), $20\pi$ (blue) and $30\pi$ (red). 
    	(a),(b) Exciton occupation $C_X$,
       	(c),(d) occupation of the energetically lower dressed state $C_{\rm{low}}$, 
       	(e),(f) instantaneous energy of the upper and lower dressed state $E_{\rm{up/low}}$, 
       	(g),(h) pulse envelopes.
       	The detuning is $\Delta=1$~meV.
	    }
 \label{fig3}
\end{figure}

In the previous section the switch-on of the laser field has been taken to be instantaneous. This implies that a sudden 
transformation of the photon dressed states occurs at the beginning of the pulse.
Similarly, when switching off the laser field the photon dressed states are transformed back to the pure ground and exciton states.
To understand the importance of these 
\emph{dressing} and \emph{undressing} processes in terms of fast and high fidelity phonon-assisted state preparation,
we now look at excitations with different pulse shapes. First we compare an excitation of the two-level system with rectangular and Gaussian pulses.
Later, we approximate the Gaussian pulse using a rectangular pulse with softened edges. 

Fig.~\ref{fig3} shows the temporal evolution under excitation for the rectangular (left) and Gaussian (right) pulses including (a),(b) the exciton occupation,
(c),(d) the lower dressed state occupation, (e),(f) the instantaneous dressed state energies and (g),(h) the pulse envelopes of the electric field.
The length of the rectangular pulses is chosen as $40$~ps, which is twice the full width of half maximum (FWHM) of the Gaussian pulses, which is set to $20$~ps.
All calculations are performed for a detuning of $\Delta = 1.0$~meV and for three different pulse areas $\alpha=\int_{-\infty}^\infty f(t)\rm{d}t$ with $\alpha=10\pi$ (green curves),
$20\pi$ (blue curves), and $30\pi$ (red curves). 

Let us start with the rectangular pulse (left panels in Fig.~\ref{fig3}). The laser pulse sets in instantaneously at $t=-20$~ps and besides damped Rabi oscillations
there is an overall increase of the exciton occupation that depends on the strength of the pulse [Fig.~\ref{fig3}(a) and see also Fig.~\ref{fig2}(a)].
In the dressed state picture this behavior can be understood as follows:
When there is no laser pulse, the dressed states are equal to the bare states where for a positive detuning the ground state corresponds to the upper dressed state $E_{\rm{up}}$ and
the exciton state corresponds to the lower dressed state $E_{\rm{low}}$. Note that in this picture the energies of the photons needed for the excitation
are counted as part of the dressed state energies.
As soon as the laser pulse sets in the dressed states become a mixture of ground and exciton state [cf. Eq.~(\ref{dressed_states})] with shifted energies shown in Fig.~\ref{fig3}(e),
i.e. the bare QD states become dressed by the laser field.
Due to their overlap with the ground-state both dressed states instantly become occupied, which is reflected by Rabi oscillations in the bare states\cite{hughes:13}.
Phonons induce transitions between the dressed states \cite{lueker:12, reiter:14} and
at low temperatures transitions to the lower dressed state, that correspond to phonon emission, outweigh.
Therefore, during the pulse the lower dressed state becomes more and more occupied as it can be seen in Fig.~\ref{fig3}(c). This also means that the exciton occupation
successively approaches its stationary value $C_X^{\infty}$, which depends on the exciton contribution to the lower dressed state but is above $0.5$ for all field strengths 
and positive detunings [c.f. Eq.~(\ref{CX_infty})].
When the laser pulse is switched off at $t=40$~ps the exciton occupation [Fig.~\ref{fig3}(a)] keeps the value it has right before the switch-off.  
At the same time the dressed states are abruptly transformed back to the bare QD states due to the sudden stop of the rectangular pulse, which means that 
the undressing takes place instantaneously. This is reflected by a step-like drop of the occupation of the lower dressed state that can be seen in Fig.~\ref{fig3}(c).
For the weakest pulse ($\alpha=10\pi$) the final state is not reached within the time window of $40$~ps,
while for the higher pulse areas the relaxation process is mostly completed. Either way, the final exciton occupation stays below $0.9$ for all of the three rectangular pulses and
in fact no matter which rectangular shape is assumed for pulse lengths of a few tens of ps one never achieves a final value close to one due to the trade-off
situation described in the previous section.

The situation is quite different for the Gaussian pulses shown in the right column of Fig.~\ref{fig3}.
In this case, no oscillations of the exciton occupation are visible and instead $C_X$ smoothly rises to its final value that for pulse areas
$\alpha=20\pi$ and $30\pi$ is considerably higher than for the rectangular pulses and practically reaches $1.0$ as it can be seen in Fig.~\ref{fig3}(b).
Like for the rectangular pulses, a phonon relaxation process takes place from the upper to the lower dressed state yielding a steady increase of the
occupation of the lower dressed state in Fig.~\ref{fig3}(d). However, because the field strength for Gaussian pulses is time-dependent,
the bare QD state contributions of the dressed states change during the pulse. This is also reflected by the time dependence of the dressed state energies shown in
Fig.~\ref{fig3}(f), which can be used to extract the Rabi splitting at a given time. 
For the weakest Gaussian pulse shown here ($\alpha=10\pi$), and similar to the case of the weakest rectangular pulse, the phonon-induced relaxation is 
too weak to complete the relaxation process within the duration of the pulse. This is because the Rabi splitting is well below the maximum of the phonon density
at energy $E_J^{\rm max}\approx2$~meV [c.f. Fig.~\ref{fig1}(c)] even at the maximum of the pulse where the Rabi splitting reaches its maximum value of approx.
$1.4$~meV [c.f. Fig.~\ref{fig3}(f)].
In contrast, for the higher pulse areas $\alpha=20\pi$ and $\alpha=30\pi$ the relaxation is more effective and leads to a full occupation of the lower dressed state.
At the pulse maximum the Rabi splitting for the $30\pi$ pulse already becomes larger ($3.0$~meV) than $E_J^{\rm max}$ temporarily leading to 
a weakening of the relaxation which is visible by a reduced gain of the lower dressed state occupation [red curve in Fig.~\ref{fig3}(d)] around $t=0$~ps.
Most interestingly, the phonon-induced relaxation is practically complete around $t=10$~ps while the exciton occupation obtained by the stronger pulses is still
far from its final value and therefore the remaining increase of $C_X$ cannot be attributed to the relaxation.
The final value of $C_X$ is only reached within a second phase of increase that takes place while the pulse is switched-off. 
This is only possible because during the switch-off the dressed states are \emph{adiabatically undressed}, i.e. adiabatically transformed back to the bare QD states.
In this process the ground state component of the then almost fully occupied lower dressed state is reduced to zero which yields
a drastic increase of the exciton occupation. Importantly, a necessary precondition for this increase 
is that the undressing takes place slow enough such that the system can follow the transformation of the dressed states adiabatically and
the occupation stays in the lower dressed state. For example this is not fulfilled for the instantaneous undressing in case of the rectangular
pulses discussed earlier where the lower dressed state occupation experiences a step-like drop at the end of the pulse [Fig.~\ref{fig3}(c)]
that does not occur for Gaussian pulses [Fig.~\ref{fig3}(d)].

As it turns out the adiabatic undressing towards the end of the pulse is in fact essential for a successful {\em fast} exciton preparation. 
The absence of the adiabatic ending prevents the phonon-assisted state preparation protocol from working with high fidelity for short 
rectangular pulses and any other pulse shapes with too fast switch-off times.

\begin{figure}[t]
 \includegraphics[width=8.7cm]{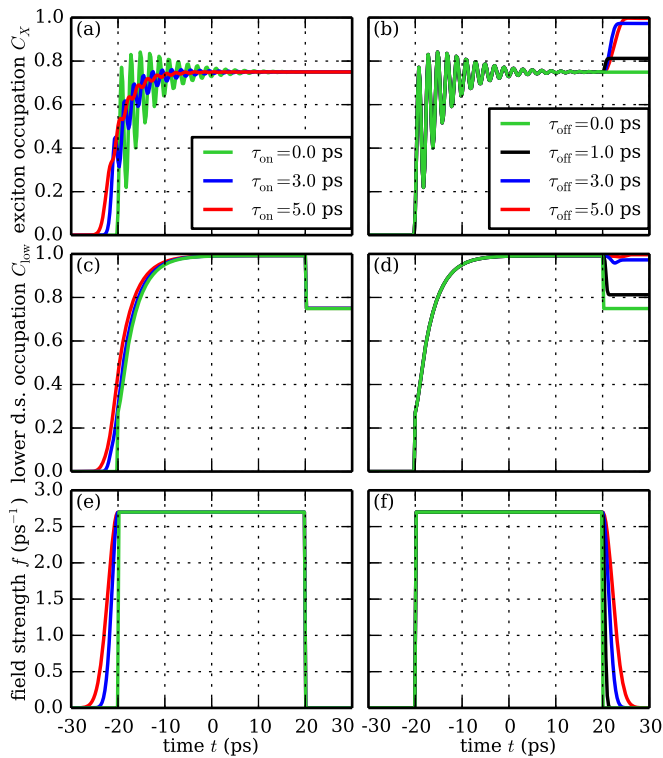}
    \caption{(Color online)
            Time dependent exciton occupation $C_X$ under pulsed excitation detuned by $\Delta=1.0$~meV for different (a) 
	    switch-on (b) and switch-off times (see key). (c),(d) corresponding
	    occupations of the energetically lower dressed state $C_{\rm{low}}$ and  (e), (f) corresponding pulse envelopes. 
	    	}
 \label{fig4}
\end{figure}

Having seen that the temporal evolution of the dressing and undressing plays a crucial role, we investigate in more detail the role of the switch-on and switch-off phase.
To this end we choose a special pulse shape that is designed to highlight the important role of these two processes on the preparation process. 
The thermalization of the dressed states is obviously achieved best by a constant part of the pulse that is chosen such that the
coupling to the phonon environment is maximal ($f = 2.7$~ps$^{-1}$ for the parameters chosen here) and that is sufficiently long
to complete the thermalization. In our case, the duration of this part of constant driving is chosen to be $40$~ps.
In order to systematically study the influences of the switch-on (switch-off) characteristics we add the left (right)
half of a Gaussian pulse with FWHM $\tau_{\rm{on}}$ ($\tau_{\rm{off}}$) before (after) the constant part of the pulse envelope as illustrated in Fig.~\ref{fig4}(e)[(f)]. 
Here, we compare a strictly rectangular pulse with $\tau_{\rm{on/off}} = 0$~ps (green) with pulses of different switch-on/off times 
$\tau_{\rm{on/off}} =1$~ps (black), $3$~ps (blue) and $5$~ps (red).
The first row [panels (a) and (b)] shows the corresponding evolution of $C_X$, while the lower dressed state occupation is plotted in the second row [panels (c) and (d)] 
and the pulse envelopes are shown in the third row [panels (e) and (f)].

Looking at the left panels of Fig.~\ref{fig4}, we find that a longer switch-on time does not alter the resulting exciton occupation after the pulse,
which is about $0.75$ for all three cases. However, we find  a significant reduction of the Rabi oscillations, which almost disappear for $\tau_{\rm{on}} = 5$~ps. 
The insensitivity of the final occupation with respect to the switch-on dynamics is related to the irreversible nature of the relaxation process which,
in case of a complete relaxation, leads to a complete loss of memory of the initial state. We thus find that the first phase, i.e., the dressing,
is necessary to start the relaxation process, but for a sufficiently long relaxation phase the details of the switch-on are irrelevant for the final fidelity. 

In contrast, as demonstrated in the right panels of Fig.~\ref{fig4}, a slower switch-off gives rise to a further gain of exciton occupation
when the field strength is reduced to zero. This effect is considerably less pronounced if the switch-off time becomes too short and does not occur at all
if the pulse stops instantaneously. For example for $\tau_{\rm{off}} = 5$~ps (red curve) $C_X$ increases by more than $0.2$ during the switch-off, 
while for $\tau_{\rm{off}} = 1$~ps (black) the increase is below $0.05$.
Thus, we conclude, that to yield a high fidelity state preparation the transformation of the dressed states into the bare QD states needs to happen 
slow enough in order to allow for an adiabatic evolution, i.e., the undressing needs to be adiabatic.

To sum up, for an efficient and fast phonon-assisted state preparation there are basically two features of the pulse envelope that are significant:
First, there has to be a phase of the pulse where the QD thermalizes in the dressed state basis, which must be long enough to complete the relaxation.
This phase can be chosen the shorter the stronger the phonon-induced relaxation is and therefore can be minimized when the field strength is such that
the strength of the phonon-coupling is maximal. Therefore in this phase a constant field strength corresponding to the maximal relaxation rate is optimal
to achieve a complete thermalization with a minimal pulse length. Second, the pulse must be switched-off slowly enough to allow the QD system for 
an adiabatic undressing of the states.

Although a Gaussian pulse shape does not meet these requirements in an optimal fashion it still works surprisingly well,
mostly because the optimum for the ideal field strength at which
the dressed state relaxation is maximal is not a sharp maximum, but rather a broad peak as it can be seen from Fig.~\ref{fig1}.
Therefore, even though a Gaussian
pulse does not have a constant plateau, a sufficient relaxation takes place during most of the time of the pulse
provided a reasonable value of the pulse area is chosen. Most importantly, deviations from the Gaussian pulse shape
that go towards faster switch-off times, as they might be induced by a pulse-shaping setup, can be harmful to 
the efficiency of the phonon-assisted preparation protocol.

\subsection{Interpretation on the Bloch sphere}\label{bloch_sphere}

\begin{figure}[t]
 \includegraphics[width=\columnwidth]{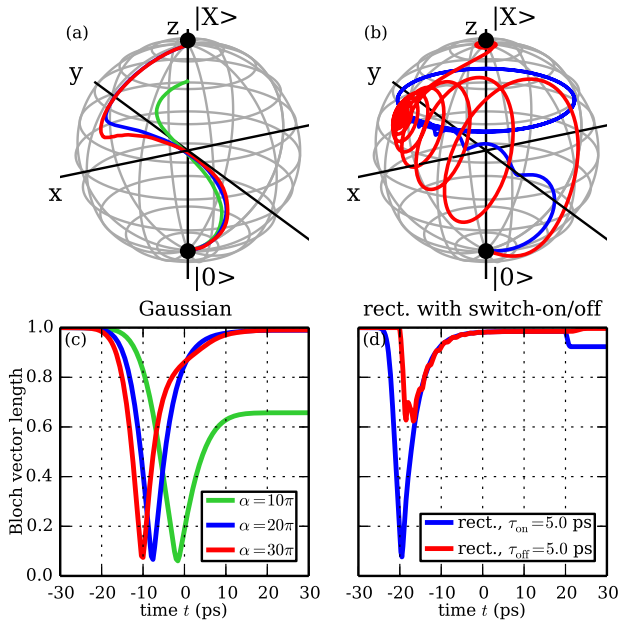}
    \caption{(Color online)
            (a),(b) Illustration of the system trajectory on the Bloch sphere and
            (c),(d) length of the Bloch vector as a function of time for 
            (a),(c) a Gaussian excitation with pulse areas $\alpha=10\pi$ (green), $20\pi$ (blue) and $30\pi$ (red)
            and (b),(d) a rectangular excitation with different switch-on/off times
            $\tau_{\rm{on}}=5$~ps; $\tau_{\rm{off}}=0$~ps (blue) and  $\tau_{\rm{on}}=0$~ps;  $\tau_{\rm{off}}=5$~ps (red). 
            The detuning is $\Delta=1$~meV.
	    }
 \label{fig5}
\end{figure}

Another interesting aspect not highlighted so far is that the incoherent phonon scattering can result in a pure state, which even 
can be transformed to a bare QD state.
This is best illustrated using the Bloch vector picture\cite{allen:75}. In this picture the projection of the Bloch vector on the $z$-axis represents the inversion,
i.e., the difference between the occupations of the upper and the lower level of the two-level system, while the in-plane component reflects the polarization.
Resonant lossless driving of a two-level system is reflected by the Bloch vector moving at the surface of the Bloch sphere taking the shortest path from one pole to the
other corresponding to the well known coherent Rabi oscillations.
Preparing the exciton state using a $\pi$ pulse then means going from the lower pole corresponding to the ground level to the upper pole 
corresponding to the excited level.
A detuned excitation leads to a tilted oscillation, which starting from the lower pole does not reach the upper pole.
If the system is fully coherent, the Bloch vector stays on the surface, i.e., its length is constantly one. If decoherence takes place,
the length of the Bloch vector is decreased. 
Clearly, the phonon-assisted preparation involves an incoherent phonon-induced relaxation, but as we have seen it can still 
eventually lead to an almost perfect exciton state.

To analyze this in more detail we have calculated the trajectory of the Bloch vector of the driven QD system, which is shown in Fig.~\ref{fig5}(a),(b)
alongside the time evolution of the length of the Bloch vector presented in Fig.~\ref{fig5} (c),(d) for an excitation with Gaussian pulses (left panel)
and for rectangular pulses with softened switch-on or switch-off (right panel). The corresponding pulse shapes can be seen in Fig.~\ref{fig3} (h) and Fig.~\ref{fig4} (e),(f).

Let us first look at the Gaussian pulses in the left column.
When the pulse is switched on the Bloch vector leaves the surface of the sphere for all pulse areas in Fig.~\ref{fig5}(a) 
[$\alpha=10\pi$ (green), $\alpha=20\pi$ (blue) and $\alpha=30\pi$ (red)]. The corresponding vector length
shown in Fig.~\ref{fig5}(c) decreases 
as it is expected as the result of the pure-dephasing phonon coupling. 
Indeed, the loss of coherence turns out to be as high as $95\%$. 
The time to reach the minimal Bloch vector length depends on the pulse area and is shorter for the stronger pulses where the relaxation is more
efficient. 
The vector length would vanish at the minimum in the ideal case of a completely adiabatic dressing process
and provided that the phonons realize a fully incoherent occupation transfer between the dressed states. In that case the electronic density matrix would
stay diagonal in the dressed state basis all the time and coherences between the dressed states would neither build up due to the laser driving nor 
due to the phonons. The continuous occupation transfer from the upper to the lower dressed state would then necessarily lead to a zero of the Bloch vector length
at some point in time. In our case, however, the minimal Bloch vector length has a small but finite value because the Gaussian pulses used in the simulations already
induce some small coherences between the dressed states (not shown).
Subsequently, the coherence between $|0\rangle$ and $|X\rangle$ is restored to a large degree and the Bloch vector length approaches $1$ for the two stronger
pulses as the phonon-assisted transitions result in an almost complete occupation of the lower dressed state which is again a pure state lying 
on the surface of the Bloch sphere. For the weaker pulse, $\alpha=10\pi$ (green), the trajectory ends inside the Bloch sphere and the vector
length stays well below $0.5$, because due to the insufficient pulse strength the relaxation does not complete.
For the stronger pulses, the dressed states transform from a superposition of ground and exciton state into
the pure exciton state during the adiabatic undressing, which corresponds to a motion of the Bloch vector on the surface towards the upper pole. 
It thus becomes obvious that the laser-driven QD evolution describes a way {\em through} the Bloch sphere to the upper pole, which is very
different to the motion along the surface in the case of an inversion yielded by applying a resonant $\pi$ pulse.

A similar analysis can be done for the Bloch vector trajectory for rectangular pulses where the switch-on or the switch-off
edge is softened. The red curve in Fig.~\ref{fig5}(b),(d) has a sharp switch-on and a smooth switch-off with $\tau_{\rm{off}}=5$~ps,
while the blue curve corresponds  to a smooth switch-on with $\tau_{\rm{on}}=5$~ps and a sharp switch-off. 
When the switch-on is instantaneous (red curve), we see that the Bloch vector trajectory exhibits a spiral movement within the Bloch sphere 
reflecting the damped Rabi oscillations. In agreement with the essentially complete relaxation, the spiral ends up on a point close to the 
surface of the Bloch sphere. For $\tau_{\rm{on}}=5$~ps (blue curve), the oscillations are less pronounced and the motion goes roughly 
along the axis of the spiral movement. The spiraling will eventually vanish for even larger values of $\tau_{\rm{on}}$, 
but the end point after the relaxation will be the same. If the laser pulse is switched off rapidly (blue curve), 
the z-component of the Bloch vector stays constant after the end of the pulse and performs a circular motion.
In contrast, for $\tau_{\rm{off}}=5$~ps an adiabatic undressing takes place during the switch-off and the Bloch vector approaches the upper pole.

\subsection{Selective state preparation in the exciton-biexciton system}\label{exciton_biexciton_system}

\begin{figure}[ttt]
 \includegraphics[width=8.7cm]{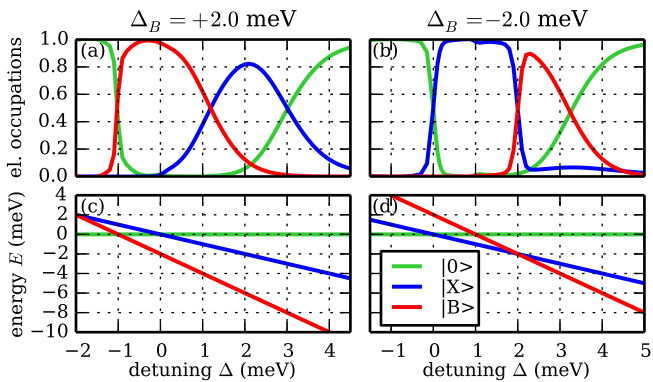}
    \caption{(Color online)
            Occupations of the electronic levels $|0\rangle$ (green), $|X\rangle$ (blue) and $|B\rangle$ (red) after optically exciting the
            exciton-biexciton system with a Gaussian pulse of pulse area $\alpha=20\pi$ and FWHM$=20$~ps as a function of the detuning $\Delta$
            for a (a) positive biexciton binding energy $\Delta_B=2.0$~meV and (b) negative biexciton binding energy $\Delta_B=-2.0$~meV. (c) and (d) :
            corresponding energies of the QD states in the rotating frame.
	    }
 \label{fig6}
\end{figure}

The clear understanding of the different phases of the phonon-assisted relaxation mechanism worked out so far turns out to be highly valuable 
to predict in an easy way the QD state that can be obtained by pulsed off-resonant excitation for arbitrary initial states and also for the 
exciton-biexciton system illustrated in Fig.~\ref{fig1}(b).
To this end it is required as before that during the switch-off phase of the pulse the system evolves adiabatically.
When the pulse duration is sufficiently long such that the relaxation all the way to the lowest
dressed state is fully completed the prepared state at the end of the pulse is determined exclusively by
the energetic order of the dressed states in the limit of vanishing pulse strength. 
Recalling that the dressed states are defined with respect to the rotating frame, the prepared state can technically be determined by subtracting 
from the energies of the QD states without light coupling the energy of the corresponding number of photons needed 
for reaching that state and looking at the resulting order of the states. More specifically, zero photons have to be
subtracted for the ground state, one photon for an exciton state and two photons for the biexciton state.
The resulting state lowest in energy will be occupied predominantly at the end of the preparation process.

For example in the two-level system subtracting the laser energy from the laser-free QD state energies yields 
dressed state energies that are in the limit of vanishing field strength separated by the detuning between the laser
and the polaron-shifted QD transition. For a positive detuning the energy of the exciton-like dressed state will be below
the corresponding ground state-like dressed state energy in the rotating frame and consequently the phonon-assisted preparation protocol with adiabatic undressing prepares 
the exciton state.  On the other hand, a negative detuning reverses the order of the dressed state energies and at low temperatures where there is mostly phonon
emission the protocol prepares the QD ground state independent of the initial state\cite{glaessl:13b}.

In the case of an exciton-biexciton system subtracting the energy of the corresponding photons gives the energies of the QD states in the frame rotating
with the laser frequency as $E'_0=0$, $E'_X = -\Delta$, $E'_B = -\Delta_B - 2\Delta$. Thus, the energetic ordering depends also on the 
biexciton binding energy and we will consider both cases with positive and negative biexciton binding energy, which can both be realized 
depending on the QD geometry\cite{michler:09}. In the following we will refer to the dressed state that in the limit of vanishing field
strength transforms into the $|0\rangle$ [$|X\rangle$, $|B\rangle$] state as $|0'\rangle$ [$|X'\rangle$, $|B'\rangle$]. 

Let us first discuss the most commonly encountered situation of a QD with a positive biexciton binding energy $\Delta_B = 2$~meV.
Figure~\ref{fig6} (a) shows the final occupation after excitation with a detuned Gaussian
pulse with a FWHM of $20$~ps and a pulse area of $\alpha=20\pi$ as a function of the detuning, while Fig.~\ref{fig6} (c) shows the corresponding energies
in the rotating frame. It can be seen that for detunings below the two-photon resonance, i.e., $\Delta < -\Delta_B/2 = -1$~meV, 
the occupation remains in the ground state (green curve). This is consistent with the energetic order of the states, since $E'_0$ is the lowest
energy in this parameter region. 
At $\Delta = -1$~meV the energetic order of the dressed states changes, because for detunings above the two-photon resonance $E'_B$ is 
the lowest energy. This leads to a significant drop of the final ground state occupation in favor of the biexciton occupation (red curve), 
which approaches its maximum close to one at $\Delta\approx-0.5$~meV. At the one-photon resonance at $\Delta = 0$, 
the energetic order changes once again and for all positive detunings $|0'\rangle$ is the highest energy dressed state, while $E'_B$ remains 
being the lowest energy. For sufficiently long pulses all the occupation would, of course, end up in the lowest branch resulting in
the preparation of the biexciton. However, for $\Delta > 1$~meV the energetic splitting $E'_X - E'_B > 3$~meV exceeds the maximum of
the phonon spectral density $E_J^{\rm max}$ lying around $2$~meV by far, which results in a very inefficient relaxation to the $|B'\rangle$ state,
which is not completed in the time window set by the pulse length.
Therefore, instead, we observe a gain of the exciton occupation (blue curve) as soon as $E'_X$ crosses  $E'_0$.
The maximal exciton occupation of about $0.8$ is reached around $\Delta = 2$~meV, where the energy splitting $E'_0 - E'_X = 2$~meV 
agrees with $E_J^{\rm max}$.
Therefore, it turns out that for a system with more than two states like the one considered here an incomplete relaxation
can also be advantageous for preparation purposes if a preparation of a QD state that in the rotating frame
is not the lowest lying state, like in our case the exciton, is intended.
An even higher exciton occupation is possible for a larger biexciton binding energy which favors transitions to 
$|X'\rangle$, because the coupling to the lowest dressed state $|B'\rangle$ in this case gets even more
out of resonance. 
It also follows that whether the prepared state is 
the exciton or the biexciton can in principle be selected by suitably adjusting the pulse length for all
positive detunings, because the final state depends on whether the relaxation completes during the pulse 
or whether the intermediate level corresponding to the exciton 
is still  predominantly occupied when the pulse is switched off. 
Detunings higher than $2.0$~meV lead to a decrease of both the final 
exciton and the biexciton occupations, because the
phonon-induced relaxation becomes weaker and weaker as the phonon environment becomes more and more out of resonance.
Finally, for very large detunings above $\Delta=4.0$~meV the energetic splittings between the dressed states are so large that for the given pulse length practically no relaxation
takes place and the QD remains in the ground-state.

The case of a negative biexciton binding energy of $\Delta_B=-2$~meV is shown in the right column. Fig.~\ref{fig6}(b) shows 
the final occupations and the corresponding energies are plotted in Fig.~\ref{fig6}(d). Similar to the case of 
positive biexciton binding energies the system remains in the ground state up to a detuning of $0$, since $E'_0$ is the lowest energy. 
Between $\Delta = 0$ and $2$~meV  $|X'\rangle$ is the dressed state with the lowest energy. This energetic order appears
exclusively for negative biexciton binding energies, resulting in a broad region where a complete preparation of the exciton occurs.
This finding is somewhat surprising, because the two-photon resonance at $\Delta = 1$~meV also lies within this interval,
where an efficient preparation of the exciton state might be unexpected, but can easily be understood in the context described here.
At $\Delta = 2$~meV the energetically lowest state changes a second time and a sharp transition of the prepared state from the exciton to
the biexciton occurs.
However, for $\Delta > 2.5$~meV the energy splittings between the dressed states that are connected to the excitonic states and $|0'\rangle$ already become too large for 
a complete thermalization to take place during the length of the pulse.
This leads to a significant reduction of the biexciton occupation for higher detunings and above $\Delta = 5$~meV the QD does not get affected by the pulse
anymore staying in the ground-state.
Importantly, also in the case of a negative biexciton binding energy the targeted state can be derived from the energetic order shown in Fig.~\ref{fig6}(d),
demonstrating the correctness of the description found for the dynamics of the phonon-induced preparation process also in the case of the 
exciton-biexciton system.

As we have seen, to selectively address a state in an exciton-biexciton system by phonon-assisted state preparation the biexciton binding energy
needs to be considered carefully, while the detuning $\Delta$ and the pulse length can act as control parameters to choose
the targeted state.


\section{Conclusions}\label{conclusions}  

We have analyzed the dynamics of an optically driven semiconductor QD coupled to longitudinal acoustic phonons for different off-resonant
excitation conditions focusing on two main questions: a) what are the requirements for a fast state preparation using off-resonant
driving and b) how is the prepared state selected. We demonstrated that a fast high fidelity preparation process not only relies on
an efficient phonon-induced relaxation between the dot-photon dressed states, but also on a successful adiabatic undressing of the states.
To understand the influence of the adiabatic undressing we compared the exciton occupations produced by Gaussian and rectangular pulse 
shapes and systematically studied the influence of the switch-on and switch-off times. This analysis revealed that while the switch-on 
time plays a subordinate role it is crucial that the pulse is switched off slowly enough. Furthermore, we analyzed the coherence properties
during the state preparation process and revealed that the incoherent phonon scattering can also restore coherence and the system trajectory takes a way through
the Bloch sphere to the opposite pole.
Finally, we showed that the concept of phonon-assisted state preparation by adiabatic undressing 
also applies to the exciton-biexciton system, where an easy prediction of the prepared state is possible.
When the pulse is long enough to support a full relaxation the prepared state is the one that is lowest in energy in the rotating frame
in the limit of vanishing field strength.
We demonstrated that even pulses too short for a full relaxation can be used for preparing intermediate states that are not the lowest 
in energy in the rotating frame. 
Altogether, it is demonstrated that in an exciton-biexciton system the decisive parameters for selecting the prepared state 
are  the detuning, the biexciton binding energy and the pulse length.

\section{ACKNOWLEDGMENTS}

A.M.B. and V.M.A. gratefully acknowledge the financial support from Deutsche Forschungsgemeinschaft
via the Project No. AX 17/7-1. D.E.R. is thankful for financial support from the German Academic Exchange Service (DAAD) within the P.R.I.M.E. programme.

\end{document}